\documentclass[a4paper,amsfonts,amssymb,amsmath,reprint,showkeys,nofootinbib,superscriptaddress]{revtex4-1}
\usepackage[english]{babel}
\usepackage[utf8]{inputenc}
\usepackage[colorinlistoftodos,color=green!40, prependcaption]{todonotes}
\usepackage{amsthm}
\usepackage{mathtools}
\usepackage{physics}
\usepackage{xcolor}
\usepackage{graphicx}
\usepackage[left=23mm,right=13mm,top=35mm,columnsep=15pt]{geometry} 
\usepackage{adjustbox}
\usepackage{placeins}
\usepackage[T1]{fontenc}
\usepackage{lipsum}
\usepackage{csquotes}
\usepackage{siunitx}
\usepackage{caption}
\usepackage{subcaption}
\usepackage{ragged2e}
\usepackage{comment}
\usepackage[pdftitle={Article},pdfauthor={Author},colorlinks=true]{hyperref}
\begin{document}
\title{The Weinberg no-go theorem for cosmological constant and nonlocal gravity}

\author{Salvatore Capozziello}
    \email{capozziello@na.infn.it}
    \affiliation{ Dipartimento di Fisica "E. Pancini", 
Universit\`a degli Studi di Napoli ``Federico II'', Via Cinthia Edificio 6, 80126 Napoli, Italy}
    \affiliation{Scuola Superiore Meridionale, Largo San Marcellino, 10, 80138, Napoli, Italy}
    \affiliation{INFN Sezione di Napoli, Complesso Universitario di Monte Sant'Angelo, Edificio 6, Via Cintia, 80126, Napoli, Italy.}
    \author{Anupam Mazumdar}
    \email{anupam.mazumdar@rug.nl}
    \affiliation{Van Swinderen Institute for Particle Physics and Gravity, University of Groningen, 9747 AG Groningen, the Netherlands}
\author{Giuseppe Meluccio}
    \email{giuseppe.meluccio-ssm@unina.it}
    \affiliation{Scuola Superiore Meridionale, Largo San Marcellino, 10, 80138, Napoli, Italy}
    \affiliation{INFN Sezione di Napoli, Complesso Universitario di Monte Sant'Angelo, Edificio 6, Via Cintia, 80126, Napoli, Italy.}

\date{\today}

\begin{abstract}
We show how a nonlocal gravitational interaction can circumvent the Weinberg no-go theorem on  cosmological constant, which forbids the existence of any solution to the cosmological constant problem within the context of local field theories unless some fine-tuning is assumed. In particular, Infinite Derivative Gravity theories hint at a possible understanding of the cosmological constant as a nonlocal gravitational effect on very large scales. In this perspective, one can describe the observed cosmic acceleration in terms of an effective field theory without relying on the fine-tuning of parameters or  additional matter fields.
\end{abstract}

\maketitle

{\it $\star$ Introduction.} The cosmological constant problem refers to the discrepancy between the observed value of the cosmological constant as a parameter of the Standard Cosmological Model and its predicted value corresponding to the energy density of the vacuum in terms of  Quantum Field Theory \cite{weinberg:cosmological}. Several solutions to this problem have been proposed in the literature, with the most popular being those invoking the dynamics of a classical field to adjust the vacuum energy density to the observed value; this class of models is essentially dismissed by the Weinberg no-go theorem on the cosmological constant, which highlights that any such \emph{local} field theory has to resort to a fine-tuning of the value of  cosmological constant. Only recently, nonlocal gravity has been put forward as a possible way out via a model which hypothesizes the existence of an extra field that achieves the cancellation of  cosmological constant from the equations of motion through a (nonlocal) spacetime averaging procedure \cite{carroll:nonlocal,oda:manifestly,oda:cosmic}.

In this letter, we exhibit a possible solution to the cosmological constant problem based on a nonlocal gravitational interaction with matter but not with any new matter degrees of freedom. We perform this approach  by analysing the structure of a unitary subclass of nonlocal theories of gravity, the so called Infinite Derivative Gravity (IDG) theories, see \cite{conroy:generalised}. However, instead of studying IDG ultraviolet properties~\cite{Biswas:2005qr, Biswas:2011ar}, we  explore the  infrared aspects. Our main result is to show  how the Weinberg no-go theorem can be invalidated if the crucial assumption of locality is lifted. This conclusion hints at the possibility of solving the cosmological constant problem by introducing nonlocal gravitational terms in an effective field theory of gravity.

The most familiar formulation of the cosmological constant problem is simply ``Why is the vacuum energy so small?'' \cite{arkani:non-local}. This reflects the rationale of the most common approach to the problem: to introduce some dynamics, analogous to the Peccei--Quinn mechanism for the strong CP problem, that is flexible enough to adjust and cancel any value of vacuum energy density. Furthermore, given the Einstein field equations of General Relativity (with $c=1$)
\begin{equation}\label{eq:einstein}
    G_{\mu\nu}+\Lambda_\textup{eff}g_{\mu\nu}=8\pi GT_{\mu\nu},
\end{equation}
the Weinberg no-go theorem can be stated as follows \cite{weinberg:cosmological}:

\begin{quotation}
    The appearance of an effective cosmological constant makes it impossible to find any solutions of the Einstein field equations in which $g_{\mu\nu}$ is the constant Minkowski term $\eta_{\mu\nu}$. That is, the original symmetry of general covariance,	which is always broken by the appearance of any given metric $g_{\mu\nu}$, cannot, without fine-tuning, be broken in such a way as to preserve the subgroup of space-time translations.
\end{quotation}

{\it $\star$ The Weinberg no-go theorem.} To prove the inevitability of the Weinberg conclusion, i.e. that a fine-tuning of the effective cosmological constant is necessary, one can consider a hypothetical solution of the field equations \eqref{eq:einstein}  preserving the symmetry of translational invariance or, in other words, a solution with all the fields being constant. The Euler-Lagrange equations of the theory for matter and gravity must then be given respectively by
\begin{align}
	\label{eq:psi}
	\frac{\partial\mathcal{L}}{\partial\psi_i}&=0,\\
	\label{eq:g}
	\frac{\partial\mathcal{L}}{\partial g_{\mu\nu}}&=0,
\end{align}
with $i=1,\dots,N$. Therefore, there are $N$ equations for the matter fields $\psi_i$ and 6 equations for the independent components of the metric $g_{\mu\nu}$, so in total there are $N+6$ independent equations for $N+6$ unknowns. As a consequence, one might expect to be able to find a solution to the above set of differential equations without any fine-tunings. The problem is that if the Eq. \eqref{eq:psi} is satisfied, then the dependence of the Lagrangian $\mathcal{L}$ on the metric $g_{\mu\nu}$ is too simple for the Eq. \eqref{eq:g} to be satisfied too and thus it is impossible to write down a Lagrangian for the physical system. In that case, in fact, diffeomorphism invariance requires \cite{weinberg:cosmological} that
\begin{equation}
	\label{eq:det-g}
	\mathcal{L}=C\sqrt{-g},
\end{equation}
where $g$ is the determinant of $g_{\mu\nu}$ and $C$ is a constant. From the result \eqref{eq:det-g}, it  follows that there are no solutions of the Eq. \eqref{eq:g} unless for some reason the coefficient $C$ vanishes when Eq. \eqref{eq:psi} is satisfied: under these assumptions, the alleged theory cannot exist. 

In the case of General Relativity, the only constant appearing in the Lagrangian of the Einstein--Hilbert action is $C\equiv-\frac{\Lambda_\textup{eff}}{8\pi G}$, so the only way for the field equations \eqref{eq:einstein} to admit a solution with all the fields constant, including the constant Minkowski metric, is for the value of $\Lambda_\textup{eff}$ to be fine-tuned to vanish in this specific case: this conclusion seems highly unnatural, given that the value of the cosmological constant is otherwise completely unconstrained by the theory.

One of the most common approaches to solve the cosmological constant problem is to invoke the presence of some new matter fields whose dynamics can be modeled in such a way as to (exactly or almost) cancel the value of the vacuum energy density. Given a weakly coupled scalar field $\phi$ and a value $\phi_0$ for which the Lagrangian $\mathcal{L}$ of a local field theory is stationary with respect to $\phi$, if the stress-energy tensor $T$ depends on $\phi$ in such a way as to vanish for the value $\phi_0$, then $\phi$ will evolve until it reaches the equilibrium value $\phi_0$, where $T=0$, and the Einstein field equations will admit a flat spacetime solution like the Minkowski metric, providing that the definition of $T_{\mu\nu}$ this time includes the contribution $-\frac{\Lambda}{8\pi G}g_{\mu\nu}$ of the cosmological constant term. 
A weak coupling implies that the equilibrium value $\phi_0$ is very large. However, it is still possible that the field $\phi$ could have important effects because it must have a very small mass $m_\phi$, since it should be $m_\phi\lesssim10^{-12}$\,GeV -- corresponding to a field with macroscopic range, as its Compton wavelength would be $m_\phi^{-1}\gtrsim10^{-4}$\,m \cite{weinberg:cosmological}. In any case,  the Weinberg no-go theorem shows that it is impossible to construct a \emph{local} field theory with one or more scalar fields having these desired properties.

{\it $\star$ The geometric construction.} To see why this is the case, one can consider the trace of the field Eqs. \eqref{eq:einstein}, which for constant fields is
\begin{equation}
	\label{eq:trace}
	g_{\mu\nu}\frac{\partial\mathcal{L}}{\partial g_{\mu\nu}}=8\pi GT.
\end{equation}
 In a local field theory,  this trace equation is a linear combination of the field equations for the matter fields, i.e.
\begin{equation}
	\label{eq:lin-comb}
	g_{\mu\nu}\frac{\partial\mathcal{L}}{\partial g_{\mu\nu}}=\sum_{i=1}^N f_i(\psi_i)\frac{\partial\mathcal{L}}{\partial\psi_i}.
\end{equation}
As an example, let us consider  the Brans--Dicke scalar-tensor theory, where the equation of motion for the scalar field $\phi$ is proportional to the trace of the field equations, that is  $\Box\phi\propto T\propto R$. 

 Eq. \eqref{eq:lin-comb} can be restated as a symmetry condition, because, for constant fields, the Lagrangian must be invariant under the transformations \cite{weinberg:cosmological}
\begin{equation}
	\label{eq:transf}
	\delta g_{\mu\nu}=2\varepsilon g_{\mu\nu},\qquad\delta\psi_i=-\varepsilon f_i(\psi_i).
\end{equation}
Once the condition \eqref{eq:transf} is known, it follows that a solution of  Eq. \eqref{eq:psi} automatically satisfies  Eq. \eqref{eq:lin-comb} too. In order to prove the validity of this claim and justify the form of  transformations \eqref{eq:transf}, one can take into account  a simple version of the proof of Weinberg no-go theorem as follows, leading to the conclusion that it is impossible to find a solution of  field Eqs. \eqref{eq:psi} for  matter fields without fine-tuning $\mathcal{L}$. To prove this, the $N$ fields $\psi_i$ can be replaced with $N-1$ fields $\sigma_j$ (not necessarily scalars) and one scalar field $\phi$, in such a way that the symmetry transformations \eqref{eq:transf} become
\begin{equation}
	\label{eq:symmetry}
	\delta g_{\mu\nu}=2\varepsilon g_{\mu\nu},\qquad\delta\sigma_j=0,\qquad\delta\phi=-\varepsilon.
\end{equation}
These symmetry conditions ensure that, for constant fields, the Lagrangian can depend on $g_{\mu\nu}$ and $\phi$ only in the combination $e^{2\phi}g_{\mu\nu}$. As discussed before (see  Eq. \eqref{eq:det-g}), when the field equations for the $N-1$ fields $\sigma_j$ are satisfied, the Lagrangian must assume the general form
\begin{equation}
	\label{eq:lagrangian}
	\mathcal{L}=e^{4\phi}\sqrt{-g}\mathcal{L}_0(\sigma_j)
\end{equation}
with $\mathcal{L}_0(\sigma_j)$ constant. Therefore, in this case, the source of  scalar field $\phi$,in its field Eq. \eqref{eq:psi}, is the trace of  stress-energy tensor:
\begin{equation}
	\frac{\partial\mathcal{L}}{\partial\phi}=T\sqrt{-g},
\end{equation}
with
\begin{equation}
	\label{eq:t-mu-nu}
	T_{\mu\nu}\equiv g_{\mu\nu}e^{4\phi}\mathcal{L}_0(\sigma_j).
\end{equation}
If there exists a value $\phi_0$ of $\phi$ for which $\mathcal{L}$ were stationary with respect to $\phi$, then the trace of the Einstein field equations (i.e.  Eq. \eqref{eq:trace}) would automatically be satisfied at this point of field space, because of condition \eqref{eq:lin-comb}. Still, there is no such stationary field value as can be seen from  expression \eqref{eq:t-mu-nu} -- unless $\mathcal{L}_0$ is fine-tuned so that it vanishes for this stationary point of $\mathcal{L}$. In this sense, the scalar field $\phi$ can always be absorbed in a redefinition of the metric, e.g. $\hat{g}_{\mu\nu}\equiv e^{2\phi}g_{\mu\nu}$, and so it cannot help with  cosmological constant problem (for it could appear in the Lagrangian only with derivative couplings), unless some fine-tuning is considered.

{\it $\star$ The locality assumption  in field theory.} The key assumption of locality behind the Weinberg no-go theorem manifests in the form of symmetry transformations \eqref{eq:symmetry}, which lie at the heart of the proof summarized above. Such transformations can only be realized if a ``transverse'' hypersurface $S$ in field space can be defined: this $(N-1)$-dimensional hypersurface must be perpendicular to the direction of the scalar field $\phi$ in field space so that any point on $S$ is specified by the values of the other $N-1$ fields $\sigma_j$. If the field $\phi$ is the $k$-th of the $N$ matter fields, that is $\psi_k\equiv\phi$, then comparing the symmetry transformations \eqref{eq:symmetry} with the more general expression \eqref{eq:transf} shows that $f_i(\psi_i)=\delta_{ik}$. As a consequence, the Weinberg geometric construction in field space can be comprehensively stated as follows.

Let the hypersurface $S$  be specified by the equation $T(\psi_i)=0$, where $T(\psi_i)$ is any function on which $\sum_{i=1}^N f_i(\psi_i)\frac{\partial T(\psi_i)}{\partial\psi_i}$ does not vanish. Since $f_i(\psi_i)=\delta_{ik}$, the last condition can be rewritten as $\frac{\partial T(\psi_i)}{\partial\phi}\ne0$, which ensures that $T(\psi_i)$ must depend on $\phi$; in particular, the choice $T(\psi_i)=\phi$ explicitly shows that $S$ is perpendicular to the direction of $\phi$ in field space, as its analytic expression becomes $\phi=0$.

This foliation of field space, which is assumed to be valid for all possible values of  matter fields, allows one to take the $N-1$ fields $\sigma_j$ as a set of coordinates on $S$. Furthermore, in general, any point $\psi_i(\sigma_j;\phi)$ of field space can be defined as the solution of the ordinary differential equation $\frac{d\psi_i}{d\phi}=f_i(\psi_i)$ subject to the condition that, for $\phi=0$, $\psi_i$ is at the point of $S$ specified by the coordinates $\sigma_j$. Note that the differential equation $\frac{d\psi_i}{d\phi}=f_i(\psi_i)$ is correctly satisfied by the condition $f_i(\psi_i)=\delta_{ik}$ assumed for the specific transformations \eqref{eq:symmetry}.
The condition that $S$ be a transverse hypersurface ensures that, at least within a finite region of field space, any $N$-dimensional point $\psi_i$ corresponds to  one point on a trajectory of $S$, i.e. it can be mapped on the $(N-1)$-dimensional hypersurface $S$ simply by calculating its orthogonal projection given by $\phi=0$. Therefore, the whole construction can be summed up by saying that it is possible to pick at least one field, $\phi$, whose value in field space is independent of  remaining $\sigma_j$ fields so that the former can be varied as required by the symmetry transformations \eqref{eq:symmetry} while the latter can be assumed to be constant (in agreement with the expressions \eqref{eq:symmetry} and \eqref{eq:lagrangian}).

{\it $\star$ A nonlocal solution.} Nonlocality  can come into play at this point. Consider, for instance, the gravitational Lagrangian of  nonlocal IDG theories  as in  Ref. \cite{conroy:generalised}, which contains the series
\begin{equation}\label{eq:nonlocal}
	R\sum_{n=1}^\infty f_{1_{-n}}\Box^{-n}R,
\end{equation}
where  $f_{1_{-n}}$ are  dimensional constants  and $\Box^{-1}$ is the inverse of   d'Alembert operator. In IDG theories, we do not need to introduce any cosmological constant  to  dynamically explain dark energy since  nonlocal terms   provide infrared gravitational corrections. See e.g.   \cite{sasaki, deser:nonlocal1,conroy:generalised,deser:nonlocal2,arkani:non-local,barvinsky:nonlocal,nojiri:screening,capozziello:nonlocal, Branko,Nojiri,Modesto,Briscese, Bajardi1, Bajardi2}.

The main idea is  that  geometric quantities $\Box^{-n}R$ can be recast as  $N\rightarrow\infty$ scalar fields, that is
\begin{equation}
	\Box^{-n}R\equiv\phi_n,
\end{equation}
with appropriate boundary conditions chosen for the nonlocal operator $\Box^{-1}$. Therefore, the model can be considered as equivalent to an effective   scalar-tensor theory with $N\rightarrow\infty$  scalar fields non-minimally coupled to gravity. See also Ref.\cite{capozziello:nonlocal}. These auxiliary fields do not represent further degrees of freedom for  nonlocal theories at  quantum level, because there are no quanta associated with them \cite{conroy:generalised,belgacem:nonlocal}. It is worth noticing that these theories are ghost-free \cite{Buoninfante,
Efimov1, Efimov2, Briscese1,Briscese2, Pius1, Pius2, Chin} and  the fields   $\phi_n$, at any given point of spacetime, requires the knowledge  of the value of Ricci scalar $R$ throughout spacetime, as a feature for nonlocality.

Not only  the fields $\phi_n$ are nonlocal, they are also related to each other by the recurrence relation
\begin{equation}
\label{ghost}
    \phi_n=\Box^{-1}\phi_{n-1},
\end{equation}
so that it is impossible to select even only one of these fields as being independent of the others.\footnote{In this case, if one wants to consider the projection with $\phi_k=0$ of an $N$-dimensional point $\psi_i$ in field space on a hypothetical $(N-1)$-dimensional transverse hypersurface $S$, for the sake of argument, then that would automatically imply that also all the other $N-1$ fields vanish (because of the recurrence relation between them), hence \emph{all} points of the infinite-dimensional field space would be projected at the origin of $S$.} As a result, in  the present IDG models, it is impossible to write down the transformations \eqref{eq:symmetry}  because the $N\rightarrow\infty$ scalar fields cannot be varied independently.\footnote{Weinberg himself pointed out the possibility of this loophole in his original work \cite{weinberg:cosmological}.} The same conclusion holds for  IDG theories that use the generally covariant d'Alembert operator instead of its inverse.

In light of the peculiar field space structure of IDG theories, a possible solution to the cosmological constant problem  is provided by the replacement, in the gravitational Lagrangian containing the cosmological constant term, of nonlocal geometric terms like that of Eq. \eqref{eq:nonlocal}.  This outcome can ultimately be ascribed to two insights. First of all, nonlocal geometric terms automatically solve the issue with the spontaneous symmetry breaking of general covariance in the theory because they vanish for a flat spacetime solution like the Minkowski metric. Secondly, a growing amount of evidence hints at the fact that such nonlocal terms can actually give rise to the presently observed acceleration in the expansion of spacetime.  In  Ref. \cite{capozziello:nonlocal}, it was shown that  nonlocal gravitational effects of IDG are compatible with the accelerated expansion of the late Universe on very large  scales without the need to postulate the existence of  cosmological constant or extra matter fields. In other words, this result could suggest a physical interpretation of the effective cosmological constant appearing in the Einstein field Eqs. \eqref{eq:einstein} as the manifestation of the nonlocal features of  gravitational field in the far-infrared regime.

An important  remark is necessary at this point. Concerning the statement on ghost-free nonlocal theories, we have to say that   this property,  and then the unitarity of the theory,  strictly depends  on the adopted  model and it is not a property valid for  any nonlocal theory. For instance, necessary conditions for unitarity are that  quantum propagators of  nonlocal theories of gravity   have only poles with real positive residues and real masses, and that the theory must be  quantized in Euclidean space. These features   have been studied in several  papers. See e.g. Refs.  \cite{Efimov1, Efimov2, Briscese1,Briscese2, Pius1, Pius2, Chin}.

{\it $\star$ Conclusions.} In conclusion, the Weinberg no-go theorem relies on the key assumption of  the validity of locality  principle. This feature does not hold for nonlocal extensions of General Relativity, such as IDG, because  the global geometric construction  to probe locality and then consider  local Lagrangian do not work. The crucial  point  is that, for a nonlocal theory, Eqs. \eqref{eq:transf} and \eqref{eq:symmetry} cannot be written and then  the proof of the  Weinberg theorem cannot be repeated. According to this statement,  we can say that the theorem is not valid in  nonlocal theories of gravity.
 
Therefore, at least at the level of an effective field theory for  gravitational interaction, nonlocal gravity allows to derive  an effective  cosmological constant  avoiding fine-tuning issues. It is interesting to develop this insight further, either by investigating more general nonlocal theories of gravity or by combining them with adjustment mechanisms such as those exploiting spacetime averages of the gravitational effective action.
    As an example, it is worth noticing that, in Ref. \cite{RLambda}, it has been shown that maximally symmetric solutions  like $R_{\mu\nu}= \Lambda g_{\mu\nu}$ are exact solutions of the equations of motion for some specific nonlocal models. Also this can be considered as a direct proof of the violation of  Weinberg theorem and points out that the topic has to be further developed towards a more general statement.
     
Finally,   the stability of  cosmological constant against quantum corrections has to be taken into account.  For instance, the running of the cosmological constant in a six-derivative theory has been studied in  \cite{Rachwal}. There,  the complete set of renormalization group  equations, including also that for  cosmological constant, is explicitly solved. In the ultraviolet regime, the  theory is  asymptotically free and describes free gravitons in Minkowski or (anti-) de Sitter  backgrounds. Ghost-like states appear and, owing to the running, these ghosts may become tachyons. Extension of the theory, as models presented here,   may change the set of beta functions and hence be capable of overcoming this problem. In a forthcoming paper, this problem will be considered in detail.

{\it $\star$ Acknowledgements}
SC and GM acknowledge the Istituto Italiano di Fisica Nucleare (INFN) iniziative specifiche QGSKY and MOONLIGHT2.  This paper is based upon work from COST Action CA21136
Addressing observational tensions in cosmology with systematics and fundamental physics (CosmoVerse) supported by COST (European Cooperation in Science and Technology).

\end{document}